\documentclass[11pt, a4paper,british]{article}
\usepackage{jcappub}

\usepackage[T1]{fontenc}
\usepackage[latin9]{inputenc}
\usepackage{color}
\usepackage{amsmath}
\usepackage{amssymb}
\usepackage{wasysym}
\usepackage{graphics}
\usepackage{epsfig}
\usepackage{graphicx}
\usepackage{subfig}
\usepackage{bm}

\usepackage{babel}



\global\long\def\lx{\lambda_{h \phi }}

\global\long\def\he{h_{\mathrm{c}}}

\global\long\def\hei{h_{\mathrm{c}0}}

\title{Early Universe Higgs dynamics in the presence of  the Higgs--inflaton and non--minimal Higgs--gravity couplings }

\author[a]{Yohei Ema,}
\emailAdd{ema@hep-th.phys.s.u-tokyo.ac.jp}

\author[b]{Mindaugas Kar\v{c}iauskas,}
\emailAdd{mindaugas.m.karciauskas@jyu.fi}

\author[c]{Oleg Lebedev,}
\emailAdd{oleg.lebedev@helsinki.fi}

\author[c]{Marco Zatta}
\emailAdd{marco.zatta@helsinki.fi}

\affiliation[a]{Department of Physics, Faculty of Science, The University of Tokyo, Bunkyo-ku, Tokyo 113-0033, Japan}
\affiliation[b]{University of Jyvaskyla, Department of Physics, P.O.Box 35 (YFL), FI-40014\\ University of Jyv\"{a}skyl\"{a}, Finland}
\affiliation[c]{University of Helsinki and Helsinki Institute of Physics, P.O. Box 64, FI-00014, Helsinki, Finland}

\arxivnumber{} %

\abstract{
Apparent metastability of the electroweak vacuum poses a number of cosmological  questions. These concern evolution of the Higgs field to the current vacuum, and its stability during and after inflation. Higgs--inflaton and non--minimal Higgs--gravity interactions can make a crucial impact on these considerations potentially solving the problems. In this work, we allow for these couplings to be present simultaneously and study their interplay. We find that different combinations of the Higgs--inflaton and non--minimal Higgs--gravity couplings induce effective Higgs mass   during and after inflation. This crucially affects the Higgs stability considerations during preheating.
In particular,  a wide range of the couplings  leading to stable solutions   becomes allowed.
}

\begin{document}
\maketitle

\section{Introduction}

The currently favored  values of the top quark and Higgs masses imply that our Universe 
is metastable \cite{Buttazzo:2013uya,Bezrukov:2012sa,Alekhin:2012py}, while its lifetime is much longer than the present age of the Universe.
The extra minimum of the Higgs potential is much deeper than the electroweak vacuum. This 
raises pressing cosmological questions: how did the Universe evolve to the energetically disfavored state and why it remained there during inflation despite large field fluctuations. In the minimal framework that includes the Standard Model (SM) and the inflaton,
 the key to these puzzles may lie in the Higgs coupling to gravity \cite{Espinosa:2007qp}  or inflaton \cite{Lebedev:2012sy}.
 On general grounds, one expects the presence of the  following interactions
at the renormalizable level 
 \begin{eqnarray}
&& -{\cal L}_{hR } = \xi H^\dagger H \hat R \;, \\
&&  -{\cal L}_{h\phi } = \lambda_{h \phi} H^\dagger H \phi^2 + \sigma H^\dagger H \phi \,,
 \end{eqnarray}
where $\hat R$ is the Ricci scalar and $\phi$ is the inflaton. 
During inflation, 
these generate large effective mass  for the Higgs field, which can drive it to  small  
field values and stabilize it at the origin \cite{Lebedev:2012sy}. Studies 
of the Early Universe Higgs evolution without the extra interactions can be  
found in \cite{Espinosa:2015qea} and \cite{Shkerin:2015exa,Hook:2014uia,Kearney:2015vba,East:2016anr,Enqvist:2014bua,
Enqvist:2015sua}.

 Even if one suppresses the above  
terms at tree level, both of them are generated by renormalization group (RG) running.
Indeed, the RG equation for $\xi$ can be found in \cite{Buchbinder:1992rb,Bezrukov:2007ep,Herranen:2014cua}, while the Higgs--inflaton coupling is generated since successful reheating requires some inflaton coupling to the SM fields, 
which at loop level induces a Higgs--inflaton coupling \cite{Gross:2015bea}. 
We also note that Higgs--inflaton interaction is generated when one eliminates
$H^\dagger H \hat R$ from the action by going to the Einstein frame.  In general,  
we expect ${\cal L}_{hR }$ and ${\cal L}_{h\phi }$ to be equally important. 

The Higgs--gravity and Higgs--inflaton couplings can stabilize the Higgs field during inflation, yet the same couplings may have a destabilizing effect after inflation. In particular, in the preheating epoch, the inflaton oscillates around its central value thereby inducing an oscillating mass term for the Higgs field. This can lead to explosive Higgs production  through parametric, tachyonic or mixed resonance \cite{Kofman:1997yn,Felder:2000hj,Dufaux:2006ee,Enqvist:2016mqj}. The resulting Higgs field variance can exceed the barrier separating the two vacua thus leading to vacuum destabilization. The consequent constraints on the couplings were derived in 
\cite{Herranen:2015ima,Ema:2016kpf,Enqvist:2016mqj}. If only the non--minimal Higgs--gravity coupling is present, one requires $\xi \lesssim 10$ \cite{Ema:2016kpf}.
This assumes the inflationary Hubble rate $H \sim 10^{14}$ GeV and the SM instability scale of order $10^{10}$ GeV, which 
we take as representative values for our analysis. On the other hand, if only the Higgs--inflaton couplings are present, the constraints are roughly $\lambda_{h \phi} \lesssim   10^{-8}$, $\vert\sigma\vert \lesssim 10^{8}$ GeV \cite{Enqvist:2016mqj}. The bounds are quite tight which restricts the range of the couplings consistent with vacuum stability  $both$ during and after inflation
to a small window. 

As mentioned above, one generally expects ${\cal L}_{hR }$ and ${\cal L}_{h\phi }$
to be equally important.
Thus, in this work, we allow for the Higgs--gravity and Higgs--inflaton couplings to be present simultaneously. We find that different combinations of the couplings are responsible for stabilizing the Higgs during inflation and after inflation. In this case, the resonances during preheating can be suppressed which allows for a wide range of 
$\xi$ and $\lambda_{h \phi}$ consistent with vacuum stability in the Early Universe.

The paper is organized as follows. In Section 2, we derive the scalar equations of motion in the Einstein frame. Higgs potential stability during inflation is studied  in Section 3.
The  centrepiece of our analysis is Section 4, where preheating in the presence of 
 ${\cal L}_{hR }$ and ${\cal L}_{h\phi }$ is analyzed. We conclude with Section 5.

\section{Equations of Motion in the Einstein Frame}

We start with  the Jordan frame action of the form
\begin{eqnarray}
S & = & \int\mathrm{d}^{4}x\sqrt{-\hat{g}}\left[\frac{1}{2}\left(1-\xi h^{2}\right)\hat{R}-\frac{1}{2}\partial_{\mu}\phi\partial^{\mu}\phi-\frac{1}{2}\partial_{\mu}h\partial^{\mu}h-\hat{V}\left(\phi,h\right)\right],\label{SJ}
\end{eqnarray}
where the hats serve  to distinguish the Jordan frame  quantities 
from those in the Einstein frame. 
In this work, we set $\hbar=c=M_{\rm Pl}=\left(8\pi G\right)^{-1/2}=1$.
The potential $\hat{V}$   is given by\footnote{Note the difference in the definition of
$\lx$ compared to that of \cite{Enqvist:2016mqj}.} 
\begin{equation}
\hat{V}\left(\phi,\chi\right)=\frac{1}{2}m^{2}\phi^{2}+\frac{1}{2}\lx\phi^{2}h^{2}+\frac{1}{2}\sigma\phi h^{2}+\frac{1}{4}\lambda h^{4}.
\end{equation}
Here $\phi$   is the inflaton and $h$ is the Higgs field in the unitary gauge,
$H=(0, v+h)^T/\sqrt{2}$. In this work, we focus on  the chaotic $\phi^2$ inflation \cite{Linde:1983gd}  for
definiteness, although our results apply more generally. The above potential includes
the most general Higgs--inflaton interaction terms at the renormalizable level. 
In what follows, we will use the step function approximation for the running Higgs quartic coupling,
\begin{equation}
\lambda (h)\simeq 0.01 \times {\rm sign} \left(h_c^{\rm SM} - \sqrt{ \langle h^2 \rangle}\right) \;,
\end{equation}
where we take for definiteness $h_c^{\rm SM} \sim 10^{10}$ GeV as the critical scale 
at which $\lambda$ flips its sign.

The non--minimal Higgs coupling to gravity \cite{Chernikov:1968zm}  can be eliminated by a conformal transformation. That is,  the action in  the Einstein frame 
is obtained via the conformal transformation of the metric, 
\begin{equation}
g^{\mu\nu}=\Omega^{-1}\hat{g}^{\mu\nu},
\end{equation}
where
\begin{equation}
\Omega\left(h\right)\equiv1-\xi h^{2}.\label{Odef}
\end{equation}
This leads to  
\begin{equation}
S=\int\mathrm{d}^{4}x\sqrt{-g}\left[\frac{1}{2}R-\frac{1}{2\Omega}\partial_{\mu}\phi\partial^{\mu}\phi-\frac{1}{2}\;\frac{6\left(\xi h\right)^{2}+\Omega}{\Omega^{2}}\partial_{\mu}h\partial^{\mu}h-V\left(\phi,h\right)\right],\label{SE}
\end{equation}
where $V\left(\phi,h\right)$ is the Einstein frame potential defined
by
\begin{equation}
V\left(\phi,h\right)\equiv\frac{\hat{V}\left(\phi,h\right)}{\Omega^{2}}.\label{VE}
\end{equation}

As a result, the gravity part of the action assumes  
  the Einstein-Hilbert form but the scalar field kinetic terms become
non--canonical. One can show that the field space in eq.~(\ref{SE})
is curved which makes it impossible to canonically normalize both fields simultaneously.
One can however introduce a canonically normalized Higgs field $h_c$ via
\begin{equation}
\mathrm{d}\he\equiv\sqrt{\frac{6 \xi^2 h^{2}+\Omega}{\Omega^{2}}}\;\mathrm{d}h.\label{canh}
\end{equation}
Although it is possible to integrate eq.~(\ref{canh}) analytically,
we have no use for such an expression. Instead, we are interested in
the small $h$  regime where the difference between the Jordan and Einstein frames
is suppressed by $h^2$ (in Planck units), that is,
\begin{equation}
\left|\xi\right|h^{2}\; , \; \xi^2 h^{2} \ll1.\label{appx}
\end{equation}
In this  approximation,  
\begin{equation}
\he\simeq h\left[1+\left(\xi+\frac{1}{6}\right)\xi h^{2}\right].
\end{equation}
The Einstein frame potential then reads 
\begin{equation}
V\left(\phi,\he\right)=\frac{1}{2}m^{2}\phi^{2}+\frac{1}{2}\left(\lx+2\xi m^{2}\right)\phi^{2}\he^{2}+\frac{1}{2}\sigma\phi\he^{2}+\frac{1}{4}\lambda\he^{4}+\ldots,\label{VEcan}
\end{equation}
where the dots denote higher dimensional  operators which are small and
therefore neglected in the following analysis.

The equations of motion are found by varying the action with respect to $\phi$
and $\he$. For our purposes, it suffices to  keep  the homogeneous
parts of the inflaton and the metric, in which case we obtain  
\begin{eqnarray}
&& \ddot{\phi}+\left(3H+2\xi\he\dot{\he}\right)\dot{\phi}+\left[m^{2}+\left(\lx+2 \xi m^{2}\right)\he^{2}\right]\phi + \frac{1}{2}\sigma\he^{2}   \simeq 0,\label{EoMj}\\
&& \ddot{h}_{\mathrm{c}}+3H\dot{h}_{\mathrm{c}}-\partial_{i}\partial^{i}\he+\left[\left(\lx+2\xi m^{2}\right)\phi^{2}-\xi\dot{\phi}^{2}+\sigma\phi\right]\he+\lambda\he^{3}   \simeq 0.
\nonumber
\end{eqnarray}
Here $\partial_{i}\partial^{i} =a^{-2} \partial_{i}\partial_{i}$ with $a$ being the scale factor.  The Hubble parameter
$H$  is given by
\begin{equation}
3H^{2}\simeq\frac{1}{2}\left(1+\xi\he^{2}\right)\dot{\phi}^{2}+\frac{1}{2}\dot{h}_{\mathrm{c}}^{2}+V\left(\phi,\he\right).\label{H}
\end{equation}

\section{Vacuum Stabilization During Inflation}

The Higgs couplings to the inflaton and to gravity modify the scalar potential during inflation. If these are sufficiently large,  the Higgs field evolves to  
electroweak  values.  As a  case study, we choose the inflaton potential $m^{2}\phi^{2}/2$
 \cite{Linde:1983gd}  with $m\sim10^{-5}$ as required by the  COBE normalization of the primordial perturbations. Although this model is on the edge of the  Planck--allowed $2\sigma$ parameter
region \cite{Ade(2015)Planck-infl} (see however \cite{Enqvist:2013eua}), it captures main features of the large field inflation framework such that our conclusions apply more generally. 

\subsection{Constraints from the Higgs evolution}

Since the main thrust of our study concerns   postinflationary Higgs dynamics,
we will make certain simplifying assumptions which facilitate our analysis during inflation.   
Let us  assume that at the initial stage the scalar potential is dominated by
the inflaton contribution $m^{2}\phi^{2}/2$  and $\phi$ undergoes a slow roll to its VEV. Furthermore, we require that the Higgs potential be convex in the relevant field range so that the Higgs  evolves to smaller values. Thus,
\begin{eqnarray}
&& m^2 \gg (\lambda_{h\phi} + 2 \xi m^2)\;  h_{{\rm c}0}^2 \;, \label{inf-domination}  \\
&& (\lambda_{h\phi} + 2 \xi m^2) \; \phi_0^2 \gg \vert \lambda \vert h_{{\rm c}0}^2 \;,\nonumber
\end{eqnarray}
where $ h_{{\rm c}0}$ and $\phi_0$ are the initial values of the Higgs field and the inflaton, respectively.
We neglect the $\sigma \phi$ term during inflation (but not after) so that our stabilization mechanism is independent of the sign of $\phi$.
The above conditions constrain the Higgs values  to be small in Planck units,
\begin{eqnarray}
 \left|\xi\right|\he^{2} &\ll& 1~,    \\
 \left|\lx\right|\he^{2}  &\ll& m^{2} ~,
\label{xih2-m2f2}
\end{eqnarray}
where the first condition was imposed in the previous subsection, while the second  follows then from    (\ref{inf-domination}). We note that larger Higgs values almost up to the Planck scale can also be treated consistently \cite{Lebedev:2012sy}.

The equations of motion (\ref{EoMj})  take the form
\begin{align}
\ddot{\phi}+\left(3H+2\xi\he\dot{h}_{\mathrm{c}}\right)\dot{\phi}+\tilde{m}_{\phi}^{2}\phi & \simeq0,\label{EoMj-inf}\\
\ddot{h}_{\mathrm{c}}+3H\dot{h}_{\mathrm{c}}+\tilde{m}_{h}^{2}\he+\lambda\he^{3} & \simeq0,\label{EoMh-inf}
\end{align}
where we have introduced the effective mass terms
\begin{align}
\tilde{m}_{\phi}^{2} & \equiv\left(1+2\xi\he^{2}\right)m^{2}+\lx\he^{2},\label{meff-infl}\\
\tilde{m}_{h}^{2} & \equiv-\xi\dot{\phi}^{2}+\left(\lx+2\xi m^{2}\right)\phi^{2},\label{meff-h}
\end{align}
and neglected the Higgs spatial gradient terms which are unimportant during inflation.

We assume that initially the Hubble parameter in eq.~(\ref{H}) is
dominated by the potential term, that is, $\dot{\phi}^{2},\:\dot{h}_{\mathrm{c}}^{2}\ll V\left(\phi,\he\right)$. Slow--roll inflation is achieved if the Hubble friction term 
in the inflaton equations of motion (\ref{EoMj-inf})
dominates   $2\xi\he\dot{h}_{\mathrm{c}}$
  and the effective inflaton mass term $\tilde{m}_{\phi}^{2}$. At the same time, the Higgs 
  evolves exponentially quickly to zero if the Higgs effective mass term $\tilde{m}_{h}^{2}$ exceeds the Hubble term and the (negative)  Higgs self--interaction term. That is,  
\begin{align}
H^{2} & \gg\left(\xi\he\dot{h}_{\mathrm{c}}\right)^{2},\:\tilde{m}_{\phi}^{2} ~~~,\label{infl-cond}\\
\tilde{m}_{h}^{2} & \gg H^{2},\:\lambda\he^{2}.\label{Higgs-cond}
\end{align}
In this case, the Higgs field evolution is well described by 
\begin{equation}
\he\simeq\hei\mathrm{e}^{-\frac{3}{2}Ht}\cos\left(\tilde{m}_{h}t\right).\label{h-sol}
\end{equation}
As follows from (\ref{inf-domination}), the above conditions are satisfied in our set--up as long as $\phi^2 \gg 1 $ which is the usual large field inflation condition and 
\begin{eqnarray}
&&  \tilde{m}_{h}  \gtrsim H \;. \label{mhH}
\end{eqnarray}
The Hubble rate and $\phi$ are almost constant in this regime, and 
 \begin{align}
\tilde{m}_{\phi}^{2} & \simeq m^{2},\\
\tilde{m}_{h}^{2} & \simeq\left(\lx+2\xi m^{2}\right)\phi^{2}.\label{meff-h-m2f2}
\end{align}

\subsection{Constraint from the flatness of the inflaton potential}

As seen from eq.~(\ref{VEcan}), both the non--minimal coupling $\xi$ and the Higgs portal 
coupling contribute to the $\phi^2 \he^2$ interaction in the Einstein frame. Closing the Higgs field in the loop, this induces the following Coleman--Weinberg correction to the inflaton potential (see e.g. \cite{Lebedev:2012sy}),
\begin{equation}
\Delta V_{\rm infl} \simeq {(\lambda_{h\phi} + 2 \xi m^2)^2\over 64 \pi^2}~\phi^4 ~\ln 
{(\lambda_{h\phi} + 2 \xi m^2)\phi^2 \over m^2} \;.
\end{equation}
During the last 60 $e$--folds of inflation, this contribution should not exceed $m^2 \phi^2/2$. Thus, taking $\phi \sim 10$, one finds\footnote{These considerations of course apply as well if only the non--minimal coupling is present. The direct Higgs--inflaton coupling in the Einstein frame induces a radiative correction to the inflaton potential which leads to the constraint $\vert\xi\vert \lesssim 10^4$. This point has not been discussed in the literature. }  
\begin{equation}
\lambda_{h\phi} + 2 \xi m^2 \lesssim 10^{-6} \;.
\end{equation}
We note that the inflaton kinetic terms are Higgs--dependent, so in principle there are  further radiative corrections from the term $h^2 \partial_\mu \phi \partial^\mu \phi$.
These are however much less important as they lead to higher derivative interactions as well as   small corrections to the kinetic terms, which we neglect.

\subsection{Summary of constraints}

Fast evolution of the Higgs field to small values requires (cf.~eq.~(\ref{mhH}))
$ \lambda_{h\phi} + 2 \xi m^2 \gtrsim m^2 $. Combining this lower bound with the upper bound coming from the radiative corrections to the inflaton potential, we obtain the allowed range for the couplings, 
\begin{equation}
10^{-10}\lesssim \lx+2\xi m^{2} \lesssim 10^{-6}.\label{thebound}
\end{equation}
This is the range we will focus on in our subsequent discussion.

There are further constraints on the initial values of the Higgs and the inflaton. In particular, requiring that the Higgs potential be convex in the relevant field range
(cf.~eq.~(\ref{inf-domination})), we get 
\begin{equation}
\phi_{0} \gtrsim \sqrt{\frac{\left|\lambda\right|}{\lx+2\xi m^{2}}}\hei.\label{stab-constr}
\end{equation}
Since $\vert \lambda \vert \sim 10^{-2}$, this implies that the initial inflaton value must be between 2 and 4 orders of magnitude greater than the initial Higgs value. 
$\phi_{0}$ is also constrained by the condition that inflation last 
 about $60$ e-folds  such that  $\phi_{0}>\phi_{*}\simeq15$, where $\phi_{*}$ is the inflaton value when cosmological scales exit the horizon.

The  approximations we have employed in our analysis restrict further the initial Higgs field values  (cf.~eqs.~(\ref{appx},\ref{xih2-m2f2})),
\begin{equation}
{\left|\lx\right| \over m^2 } \hei^2~,~\vert \xi\vert \hei^2 ~,~ \xi^2 \hei^2 
~\ll~ 1~.
\end{equation}

We note that although some of the above conditions, e.g. that the inflaton dominates the energy density, can be relaxed (see \cite{Lebedev:2012sy},\cite{Kamada:2014ufa}), 
these are not essential for our analysis since our goal is to understand an interplay of
the inflationary and preheating constraints on the Higgs couplings.

\section{Higgs Evolution after Inflation: Preheating Epoch  \label{subsec:preh}}

At the end of inflation, the slow--roll conditions are violated and the
inflaton rolls rapidly to the origin where it oscillates with high
frequency. These oscillations cause  resonant particle production
of all the fields which interact strongly enough with the inflaton,
the process called ``preheating'' \cite{Kofman:1997yn,Dufaux:2006ee}.
In the case of the Higgs field, this could have disastrous consequences
since  the field can be  excited beyond the stability scale. Thus, the same
interactions which stabilize the Higgs during inflation could destabilize
it afterwards. This phenomenon was studied in detail in refs.~\cite{Herranen:2015ima,Ema:2016kpf}
for the $\lx=0$ case and in refs.~\cite{Ema:2016kpf,Enqvist:2016mqj}
for $\xi=0$.\footnote{Related analyses can be found in \cite{Kohri:2016wof,Kohri:2016qqv}.} Here we  consider a  general situation with $\lx\ne0$ and
$\xi\ne0$. We find that an interplay of the two couplings is important leading to different conclusions compared to the cases considered before. In what follows, we  drop the scale dependence of $\xi$ over the energy range relevant to inflation/preheating (see the RG equations in  \cite{Bezrukov:2007ep}) such that it should be understood approximately as $\xi (H)$, while the scale dependence of $\lambda_{h\phi}$ is completely negligible.

At the end of inflation, the Higgs field is anchored  at the origin, making
the contribution of Higgs--inflaton crossterms in eq.~(\ref{EoMj})
negligible. The Hubble friction term is also small compared to the
inflaton mass. Hence, eq.~(\ref{EoMj}) reduces to  the
  harmonic oscillator  equation with a decaying amplitude,
\begin{equation}
\phi\left(t\right)\simeq\Phi\left(t\right)\cos\left(mt\right).\label{phi-osc}
\end{equation}
Strictly speaking, the harmonic approximation above is valid somewhat
after the end of inflation. For definiteness, we can take
\begin{equation}
\Phi\left(t_{0}\right)\simeq1.
\end{equation}
Thereafter, the universe is dominated by inflaton oscillations, which
effectively behave as pressureless dust. Hence the oscillation amplitude
decays as
\begin{equation}
\Phi\left(t\right)\simeq\frac{\sqrt{8/3}}{mt},\label{Phi}
\end{equation}
while the scale factor $a$ grows as $a \propto t^{2/3}$.
Plugging eq.~(\ref{phi-osc}) into the Higgs equation of motion (\ref{EoMj}),
we get (cf. \cite{Enqvist:2016mqj})
\begin{equation}
\frac{\mathrm{d}^{2}X_{k}}{\mathrm{d}z^{2}}+\left[A_{k}\left(z\right)+2p\left(z\right)\cos 2z
+2q\left(z\right)\cos 4z  + {\delta m^2 (z) \over m^2}\right]X_{k}
\simeq0,\label{EoM-preh}
\end{equation}
where $X_{k}\equiv a^{3/2}h_{k}$ is the rescaled Fourier $k$--mode of
the Higgs field $\he$ and we have defined
\begin{align}
z & \equiv\frac{1}{2}mt,\\
A_{k}\left(z\right) & \equiv\left(2\frac{k}{am}\right)^{2}+2\left(\lx+\xi m^{2}\right)\frac{\Phi\left(z\right)^{2}}{m^{2}},\\
p\left(z\right) & \equiv 2\frac{\sigma \Phi\left(z\right)}{m^{2}},\label{p-def}\\
q\left(z\right) & \equiv\left(\lx+3\xi m^{2}\right)\frac{\Phi\left(z\right)^{2}}{m^{2}},
\label{q-def}\\
\delta m^2(z) & \equiv 3 \lambda a^{-3} \langle X ^2  \rangle . 
\end{align}
In this equation, we have resorted to  the Hartree approximation $h^3 \rightarrow 3h \langle h^2 \rangle$ such that  the Higgs self--interaction is replaced by an effective mass term. In this case, the equations for different $k$--modes decouple.
However, in our lattice simulations we do not employ this approximation. 

We note that during preheating, the $\sigma \phi h^2_{\rm c}$ term plays an important role since
it decreases slower  than the quartic $\phi^2 h^2_{\rm c}$ interaction and at some stage becomes dominant \cite{Enqvist:2016mqj}. However, the effect of this term can be isolated
and it is instructive to start with the case $\sigma=0$.

\subsection{$\sigma=0$ case: modified Mathieu equation}

Let us set $\sigma=0$ and, furthermore, neglect for the moment the Universe expansion 
and Higgs self--interaction, $\delta m^2 \rightarrow 0$. These are good approximations at the beginning of the preheating epoch. In this case, 
 eq.~(\ref{EoM-preh}) reduces to
the familiar Mathieu  equation \cite{Kofman:1997yn},
\begin{equation}
\frac{\mathrm{d}^{2}X_{k}}{\mathrm{d}z^{2}}+\left[A_{k} 
+2q \cos 4z   \right]X_{k} 
\simeq0,
\end{equation}
 In effect, the Higgs field receives an oscillating mass term which can cause   resonance. 
Certain $k$--modes up to $k_{*}/a \sim  \left(\lx+3\xi m^{2}\right)^{1/4}  \sqrt{ m \Phi} $ would undergo  a resonant growth, which results in the increase of the Higgs variance $\langle  h^2_c \rangle $,
\begin{equation}
 \langle h^2_c\rangle \simeq \int \frac{d^3 k }{(2\pi a)^3}\frac{n_k}{\omega_k}\,,
 \end{equation}
 where the mode frequency is determined by $\omega^2_k=\left(A_{k} 
+2q \cos 4z   \right)m^2/4 $ and $n_k$ is the corresponding  occupation number 
(see \cite{Enqvist:2016mqj} for further details).
 If the fluctuations become too large, the Higgs field crosses over the barrier into the catastrophic vacuum. 
 
   \begin{figure}
 \center
 \includegraphics[width=.6\textwidth]{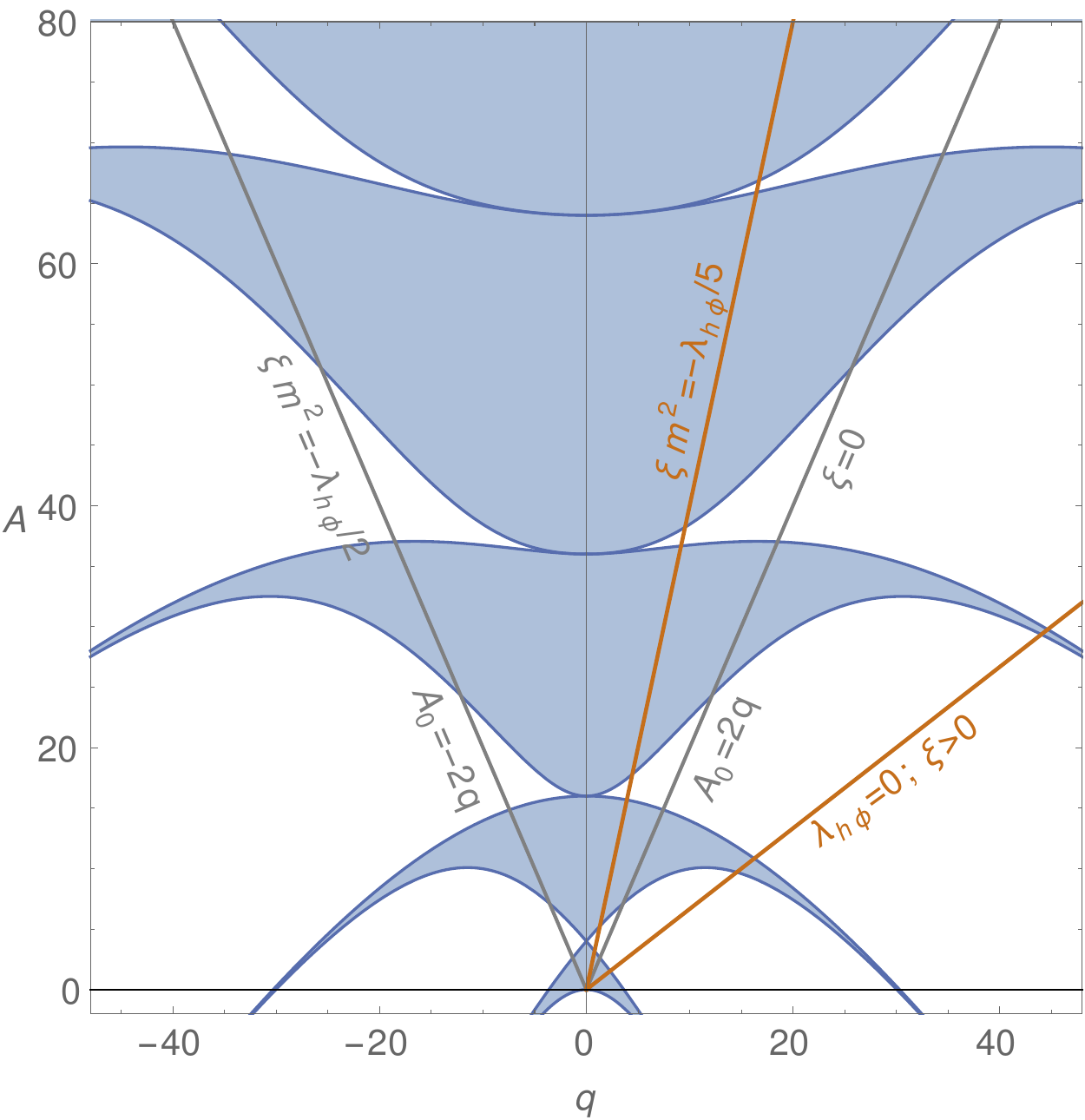}
 \caption{\label{fig1}  Stability chart of the Mathieu equation. }
\end{figure}

Behaviour of the Mathieu equation solutions is described by the stability chart 
of fig.~\ref{fig1}. If the point $(A_k,q)$ lies in the white regions, the corresponding solution grows exponentially in time.  On the other hand,
points in the shaded regions lead  to stable solutions. In an expanding Universe, both 
$A_k$ and $q$ decrease in time crossing many instability regions. The resonance ends when the system enters the last stability region at $q\lesssim4$. If the resulting $\langle  h^2_c \rangle$ is  below its critical value, no destabilization occurs.
The position of the barrier separating  the two vacua 
 is affected by the Higgs--inflaton couplings,
 \begin{equation}
 h_{\rm crit} \sim \sqrt{ 2(\lx+2\xi m^{2}) \over \vert \lambda\vert } \vert \phi\vert\;. 
 \end{equation}
Therefore even when the Higgs fluctuations exceed $10^{10}$ GeV, the system may remain stable. One finds that the destabilization criterion amounts roughly to $
\sqrt{\langle  h^2_c \rangle} >h_{\rm crit} $ with  some average value of $\vert \phi\vert$.  
 This is discussed in   detail in    \cite{Enqvist:2016mqj} 
(for $A=2 q$).

An important feature of the stability chart in fig.~\ref{fig1} is that larger $A/q$ lead
to a weaker resonance since the instability regions become shorter along the corresponding evolution lines.  In particular,   $q \simeq 0$ or 
\begin{equation}
\lx \simeq - 3\xi m^{2}
\end{equation} 
suppresses the resonance completely.  It is in fact sufficient to have $A/\vert q\vert  \sim$ few
to avoid vacuum destabilization. Such a relation is not very unnatural in our framework:
since $m^2 \sim 10^{-10}$, values of $\vert \xi \vert$ in the range 1...$10^4$ correspond to 
$\lambda_{h\phi}$ between $10^{-10}$ and $10^{-6}$. The latter is the range considered in 
 \cite{Lebedev:2012sy}.

  \begin{figure}[t!]
\begin{center}
\includegraphics[width=0.46\textwidth]{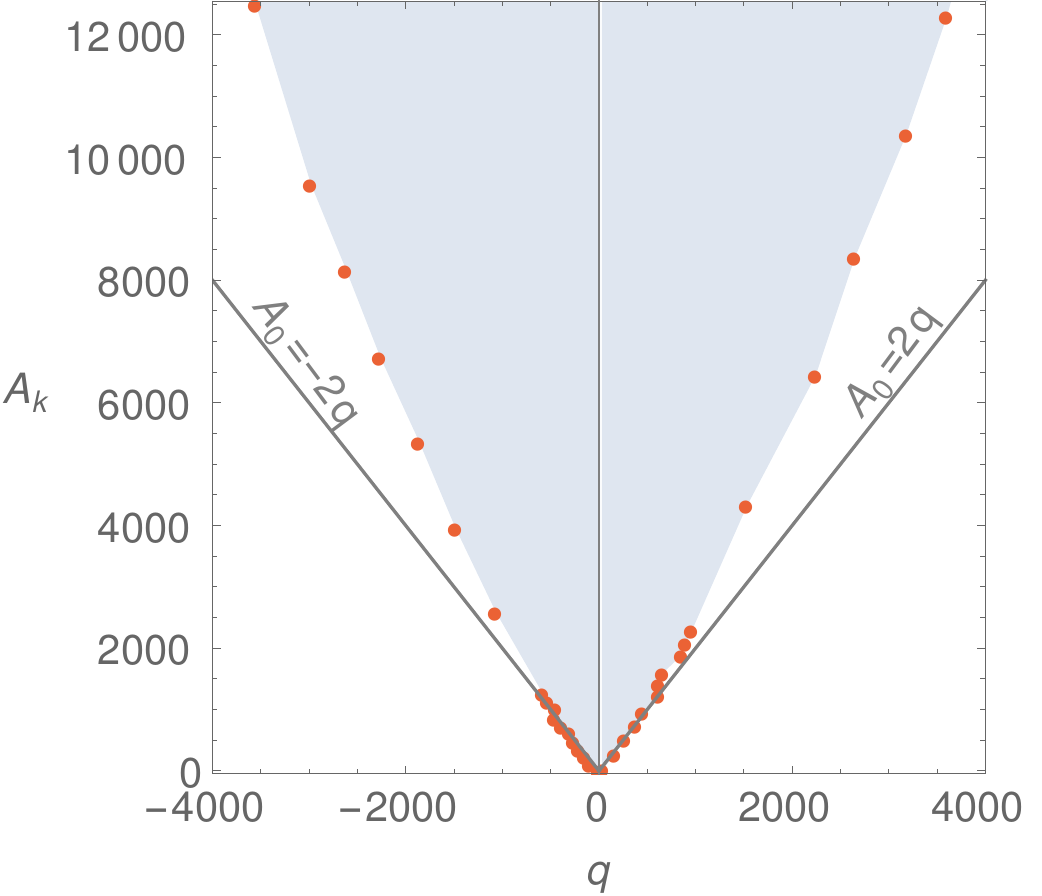} 
\includegraphics[width=0.53\textwidth]{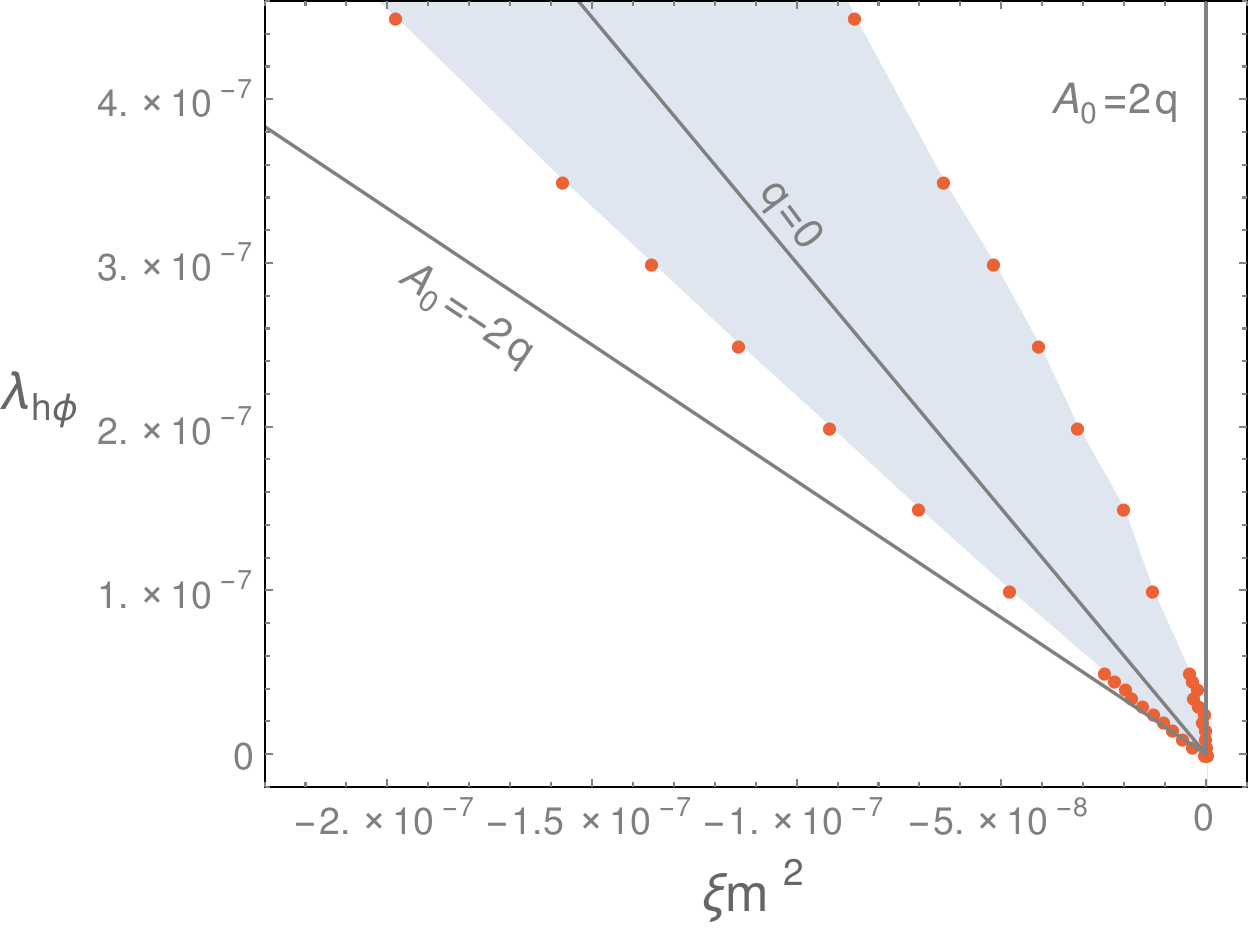} \\ 
\caption{ Vacuum stability regions (shaded) during preheating  in the $(A,q)$ plane (left) and $(\lambda_{h\phi}, \xi m^2)$ plane (right). Here $m^2\simeq 10^{-10} M_{\rm Pl}^2$ and the SM instability scale is $10^{10}$ GeV. The red dots indicate the boundary of the stability region obtained with  a modified version of \texttt{Latticeeasy}. }
\label{fig:results}
\end{center}
\end{figure}

In reality,  
the Universe expands and the Hartree approximation for the $h^4$ term 
is not necessarily reliable (see \cite{Enqvist:2016mqj}), hence the Mathieu equation gives
only an approximate description of the system behaviour. 
To determine the stability regions in our parameter space reliably,
we have resorted to classical lattice simulations (for a recent overview see \cite{Figueroa:2016wxr}).
We have used two different implementations, in the Jordan 
and Einstein frames, with two different codes:  a
modified version of \texttt{Latticeeasy}~\cite{Felder:2000hq}
and our own lattice code.
The results of our numerical
simulations are presented in fig.~\ref{fig:results}. 
These show maximal allowed $\vert q\vert $ for a given $A$ and the corresponding range
of $\xi m^2$ for a given $\lambda_{h\phi}$.
The displayed points correspond to stable configurations in the sense that the Higgs variance does not blow up before the end of the resonance. The resonance ends when the occupation numbers in the comoving frame remain constant. We find that variations of these requirements lead to similar results. Our main result is that a wide range of positive $\lambda_{h\phi}$ and negative $\xi  $ is allowed. On the other hand, negative $\lambda_{h\phi}$ lead to larger $q$ versus $A$ and therefore are ruled out (apart from small values around the origin). Note that the line $A= 2\vert q\vert$ separates the parametric resonance from the tachyonic one.

   \begin{figure}[t!]
\begin{center}
\includegraphics[width=0.45\textwidth]{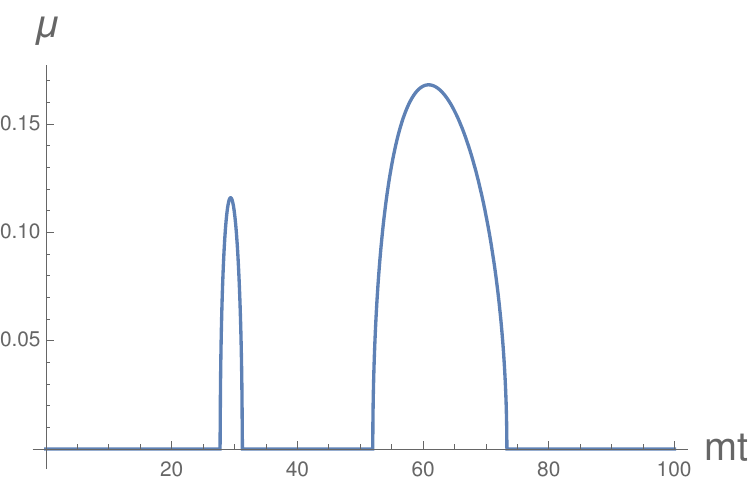} \qquad
\includegraphics[width=0.45\textwidth]{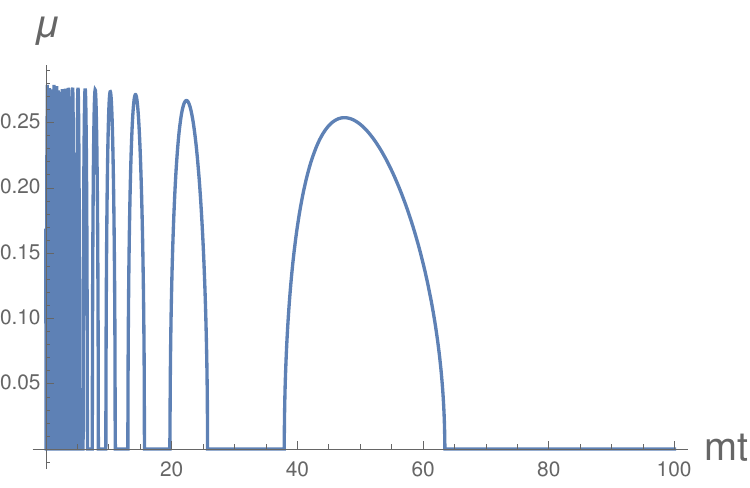} \\ 
\vspace{0.75cm}
\includegraphics[width=0.45\textwidth]{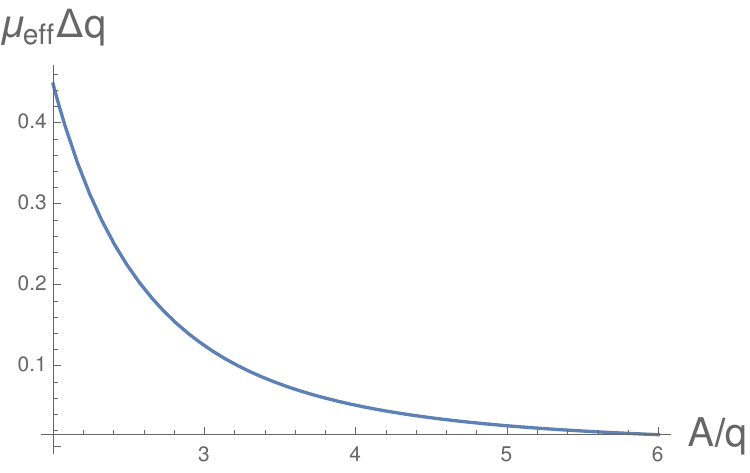}
\\
\caption{ Top: the Floquet exponent $\mu$ (cf.~eq.~(\ref{enhancement})) along the trajectory $A/q=3$ (left) and  $A/q=2$ (right). Here the initial value of $q$ is chosen to be $2000$.
 Bottom:  $\mu_{\rm eff} \Delta q$ vs
$A/q$.}
\label{fig:properties}
\end{center}
\end{figure}

The qualitative behavior of our bound can be understood as follows. 
We are mostly interested in region with a substantial ratio $A/q$. 
At $A/q > 2.3$ or so, the parametric resonance in the expanding Universe simplifies.
Whether the system gets destabilized or not is mostly determined by its behavior in the last instability region, that is,  the one closest to the origin.
This is because, in the parameter range of interest, the system spends little time in other  instability  bands since they are shorter (fig.~1)  and the inflaton evolves faster at  earlier times.  This tendency is clearly seen in fig.~\ref{fig:properties}: at $A/q=3$
only the last  band contributes significantly, whereas in the ``usual'' case of $A/q=2$ 
many bands are important.
 The increase of the Higgs amplitude for a given $k$, which can be taken $k=0$ as a representative value, is determined approximately by
 the Floquet exponent 
\begin{equation}
X_0 \propto e^{\mu_{\rm eff} \Delta mt} ,     \label{enhancement}
\end{equation}
 where $\mu_{\rm eff}$ is an average Floquet parameter $\mu$ along the relevant trajectory in the last instability band and $\Delta mt$ is the time the system spends there. Since both $A$ and $q$ have the same dependence on the inflaton amplitude, the system evolves along a straight line in the $(A,q)$ plane.\footnote{We have verified that this also applies to  the relevant range of $k \not= 0$.} 
Clearly, the resonance becomes inefficient for $ \mu_{\rm eff} \Delta mt \lesssim {\cal O}(1)$.
 The quantity  $\Delta mt$
  depends  on two factors: (a) $A/q$ determines the width $\Delta q$ (as well as $\mu_{\rm eff}$) of the instability band with larger $A$ leading to smaller $\Delta q$; (b) the rate of the inflaton change in the last band which is controlled by the
duration of the resonance $mt_{\rm end}$. 
 For reference, in fig.~\ref{fig:properties} (bottom) we display the scaling  of $\mu_{\rm eff} \Delta q$ with $A/q$. 
This can be converted to $\mu_{\rm eff} \Delta mt$ using the approximate 
relations $\Delta q/q \simeq 2 \Delta mt/mt$ and $mt_{\rm end}\simeq {\rm const }\times \sqrt{\lambda_{h\phi} +3\xi m^2} $.
We thus get the following scaling
\begin{equation}
\mu_{\rm eff} \Delta mt \simeq  {\rm const }\times \sqrt{\lambda_{h\phi} +3\xi m^2} \;  \mu_{\rm eff} \Delta q \;.
\end{equation}
This shows that the resonance can be suppressed at larger couplings
$\lambda_{h\phi} +3\xi m^2$, i.e. larger initial $q$,  by increasing $A/q$. For instance, a tenfold increase in $\lambda_{h\phi} +3\xi m^2$ can be compensated by increasing 
$A/q$ from 2 to 3. This is roughly what we observe in fig.~\ref{fig:results}.  
While for larger $A/q$ the above scaling works well,
at $A/q \leq 2.3$ this approximation breaks down and many instability bands contribute to the resonance. In fact, the tachyonic resonance is also consistent with vacuum stability as
long as the relevant couplings are small, i.e. around the origin in fig.~\ref{fig:results}.

\begin{figure}[t!]
\begin{center}
\includegraphics[width=0.8\textwidth]{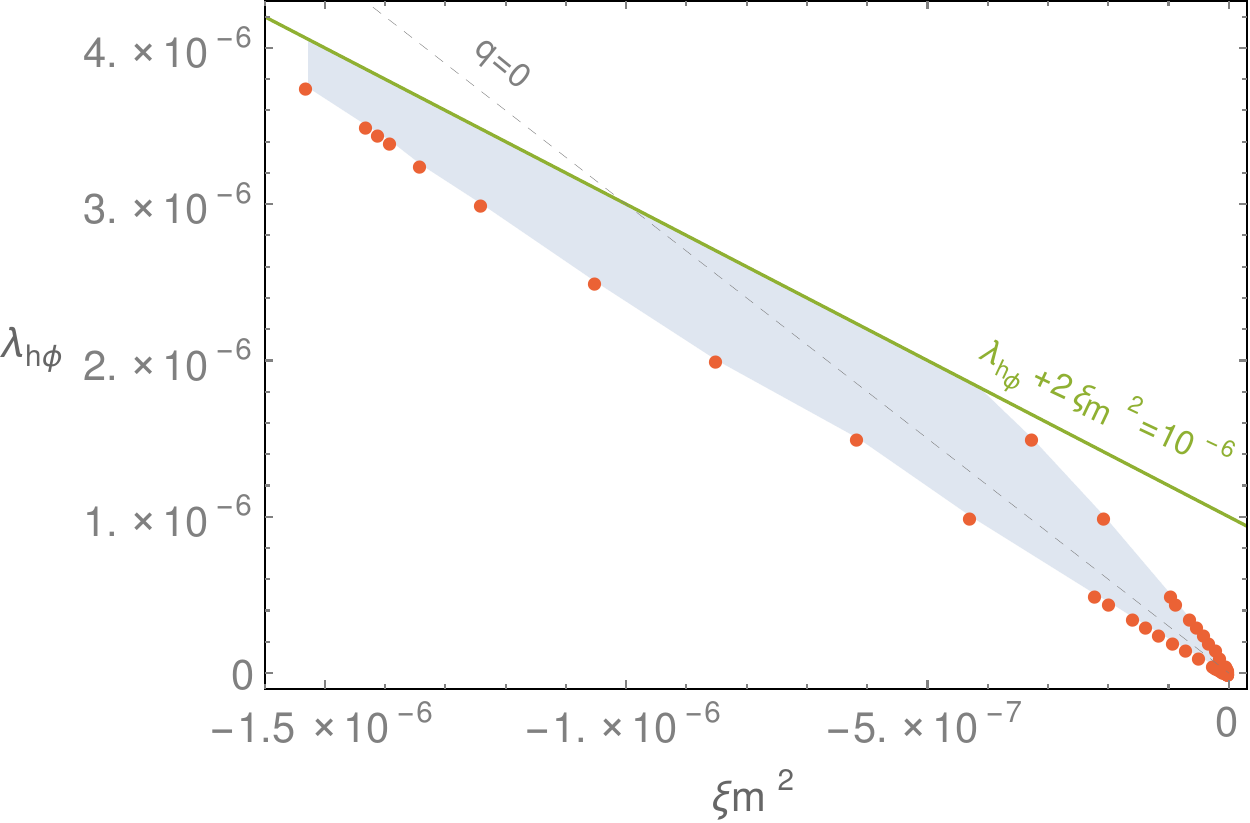} 
\caption{ Parameter space consistent with vacuum stability during inflation and preheating
(shaded). Here $m^2\simeq 10^{-10} M_{\rm Pl}^2$ and the SM instability scale is $10^{10}$ GeV. The green line marks the upper bound on the couplings from the flatness of the inflaton potential.
  The red dots indicate the boundary of the stability region obtained with  our lattice simulations. }
\label{fig:final-result}
\end{center}
\end{figure}

The final results of our study are shown in fig.~\ref{fig:final-result}, which displays
the parameter space consistent with vacuum stability during and after inflation. The region is finite due to the large couplings being cut off by the    constraint 
of sufficiently  flat inflaton potential. Negative values of $\lambda_{h \phi}$ (and positive
$\xi$) lead to   stronger   resonance and thus are excluded except for points close to the origin (as in \cite{Ema:2016kpf},\cite{Enqvist:2016mqj}).
 In the plot, the red dots indicate  the boundary of the stability region obtained with  our lattice simulations. These center around the ``no resonance'' $q=0$ line.
  The allowed parameter space extends to about $\lambda_{h \phi}\sim 6 \times 10^{-6}$, although the tip of this region is not shown  in the plot.  The inflationary  constraint  $\lambda_{h\phi} + 2 \xi m^2 > 10^{-10}$ is satisfied automatically by points in the shaded  
 region.
 
 Let us emphasize again the difference between the analyses with and without the 
 non--minimal Higgs coupling to gravity. The $\xi $--term  
   brings in both the $h^2_c \phi^2$ and $h^2_c \dot\phi^2$  interactions during preheating, thereby modifying $A_k $ and $q$ in different ways. This changes radically the stability analysis compared to the pure Higgs--inflaton coupling case since $A/q$ becomes a free variable. Hence, the resonance can be suppressed while still retaining the positive stabilizing effect of the couplings during inflation.
 
  \subsection{$\sigma\not= 0$ case: modified Whittaker--Hill  equation}

The trilinear interaction $h^2_c \phi$ can be neglected compared to $h^2_c \phi^2$ during inflation, however when
the inflaton field decreases in the preheating epoch, it can become dominant.  
While the quantities $q$ and $ A_k$ (for small $k$)  in eq.~(\ref{EoM-preh})  decrease as $1/t^2$, the coefficient $p$ decays as $1/t$ which leads to a tachyonic resonance at late times. This leads to very efficient Higgs production and thus to a strong constraint on $\sigma$ \cite{Enqvist:2016mqj}.   
Neglecting the Universe expansion and the Higgs self--interaction, the Higgs dynamics are
described by the Whittaker--Hill  equation (cf.~\cite{Whittaker(1996)preh,Lachapelle(2009)preh,Roncaratti(2010)preh,Possa(2016)preh})
\begin{equation}
\frac{\mathrm{d}^{2}X_{k}}{\mathrm{d}z^{2}}+\left[A_{k} 
+ 2p \cos 2z  + 2q \cos 4z   \right]X_{k} 
\simeq0.
\end{equation}
This is a good approximation as long as $A_k, p $ and $q$ change adiabatically. 
The relevant stability charts for the Whittaker--Hill  equation have been studied in 
\cite{Enqvist:2016mqj}. The results of that study can be directly applied here since the presence of the non--minimal coupling to gravity does not affect the trilinear coupling.

 \begin{figure}[t!]
\begin{center}
\includegraphics[width=0.493\textwidth]{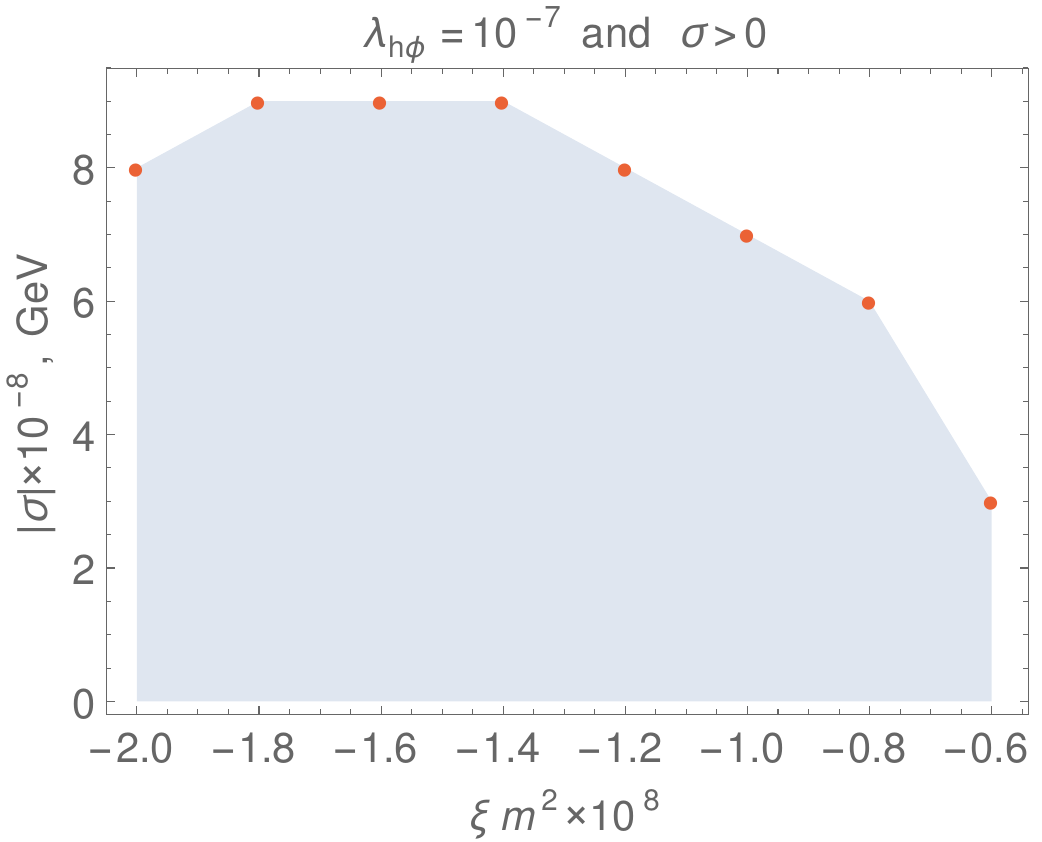} 
\includegraphics[width=0.493\textwidth]{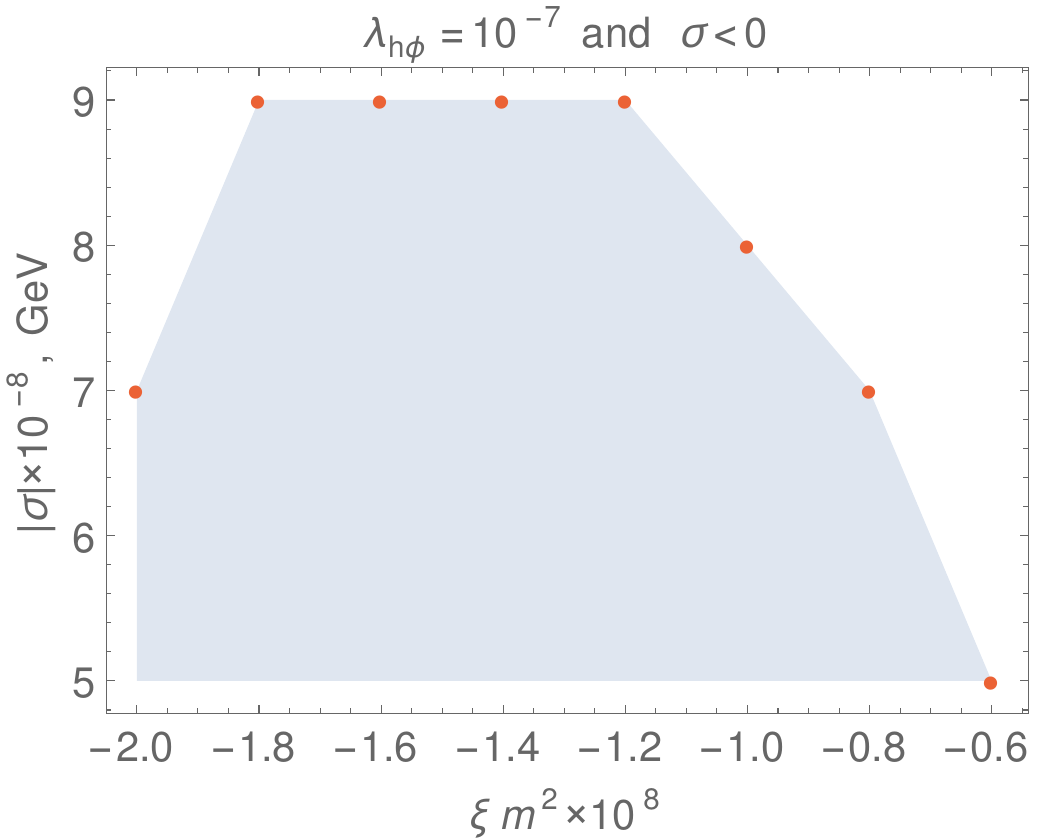} \\ 
\caption{ Vacuum stability regions (shaded) in the presence of the trilinear Higgs--inflaton coupling $\sigma$.  Left: $\sigma>0$, right: $\sigma<0$. The red dots indicate the boundary of the stability region obtained with  our lattice simulations. }
\label{fig:sigma}
\end{center}
\end{figure} 

Nevertheless, we have performed an additional  lattice study to set limits on $\sigma$.
It includes  the effects of the Universe expansion and the full Higgs self--interaction
$h^4_c$. 
Choosing $\lambda_{h\phi}$ and $\xi$ values from the stability region in fig.~\ref{fig:results}, we increase $\vert \sigma \vert$ until the tachyonic resonance destabilizes the system. Thus we determine the maximal allowed $\vert \sigma \vert$ for 
given $\lambda_{h\phi}$, $\xi$ shown in fig.~\ref{fig:sigma}. The resulting dependence is non--monotonic, yet   the order of magnitude of the maximal  $\vert \sigma \vert$ remains between $10^8$ and $10^9$
GeV, as in our previous study  \cite{Enqvist:2016mqj}. Note that positive and negative $\sigma$ are inequivalent due to the Universe expansion.

Finally, let us comment on two effects which are not taken into account in our simulations:
(a) perturbative Higgs decay and (b) possibility of late time   Higgs destabilization (after preheating). The effect of the Higgs decay into top quark pairs was studied in 
\cite{Ema:2016kpf},\cite{Enqvist:2016mqj}. It was found that although it reduces 
$\langle h^2_c \rangle$, it does not significantly affect the stability properties of a configuration with  given couplings. 
In particular, unstable configurations typically remain unstable except the destabilization time moves to somewhat larger values. 
The decay has a stabilizing effect on the system, yet it is not significant enough for our purposes and we neglect it.
Concerning the late time behavior,  $\langle h^2_c \rangle$ decreases slower in time than the position of the barrier between the two vacua \cite{Ema:2016kpf},\cite{Enqvist:2016mqj}
which can lead to eventual destabilization.
However, there are many factors that affect the system after the end of the resonance:
thermalization, non--perturbative effects, etc. Therefore, this question requires a dedicated study. Let us note however that our mechanism allows one to suppress the resonance completely such that no significant  $\langle h^2_c \rangle$ would arise. In this case, there is no danger of late time vacuum destabilization.

\section{Conclusions}

The Higgs--inflaton and Higgs--gravity couplings are essential for understanding the Early Universe Higgs dynamics. These can stabilize the electroweak vacuum during inflation  by inducing large effective  Higgs mass. On the other hand, the same couplings can have a destabilizing effect after inflation due to explosive Higgs production via parametric or tachyonic resonance. 

In this work, we have studied a combined effect of the Higgs--inflaton  $\lambda_{h\phi}$ and non--minimal 
 Higgs--gravity $\xi$ couplings. We find that a non--trivial interplay of the two allows us
 to stabilize the Higgs field during inflation while avoiding excessive Higgs production after inflation. In particular, different combinations of $\lambda_{h\phi}$ and $\xi$
 enter the $A$ and $q$ parameters of the Mathieu equation which makes it possible to suppress the parametric and tachyonic resonance. This mechanism is effective  even at ``large'' couplings up to $\lambda_{h\phi} \sim 6\times 10^{-6}$ (fig.~\ref{fig:final-result}). It requires $\lambda_{h\phi} >0$ and $\xi<0$ (unless the couplings are small, e.g. $\xi<10$).  To delineate parameter space, we have resorted to classical lattice simulations based on  \texttt{Latticeeasy} as well as an independent code.
 
 We have focused on a chaotic $m^2$ inflation model as a representative example of large field inflation and also fixed the Standard Model instability scale to be around $10^{10}$ GeV.  In this case, $\lambda_{h\phi}$ and $\xi m^2$ (where $m$ is the inflaton mass) must  be of the same order of magnitude for our mechanism to work.  Given that $m^2\sim 10^{-10}$ in Planck units, this allows for a wide range of the couplings including $\vert\xi\vert \gg 1$, unlike the previous analyses.
 
 We have also studied the effect of the mixed parametric--tachyonic resonance induced by the trilinear Higgs--inflaton coupling $\phi h^2$. The resulting bound on the trilinear parameter $\vert\sigma\vert$ is between $10^8$ and $10^9$ GeV, which is consistent with our previous study  \cite{Enqvist:2016mqj}.

\section*{Acknowledgments}

The work of Y.E. was supported in part by JSPS Research Fellowships for Young Scientists and by the Program for Leading Graduate Schools, MEXT, Japan. M.K. is supported by the Academy of Finland project 278722. 
O.L. and M.Z. acknowledge support from the Academy of Finland project ``The Higgs and the Cosmos''.

\appendix

\section{Vacuum Fluctuations and Lattice Simulations \label{appendix}}

In this appendix, we summarize the issues of vacuum fluctuations relevant to our simulations. In order   to quantify particle production,  
we need to account for vacuum fluctuations appropriately. Indeed, since the Mathieu equation is homogeneous, no particle production  occurs unless $X_k$ or its derivative is
non--zero initially. Such initial conditions are provided by the vacuum fluctuations.
In \texttt{Latticeeasy}, one employs the probability distribution for the ground state of a scalar field in a Friedmann Universe \cite{Polarski:1995jg},
 \begin{equation}
P(X_k) \propto  \exp\left(-2\omega_k^2\vert X_k\vert^2\right)\,\,,
\end{equation}
where  $\omega_k$ is the frequency of oscillations of the eigenmode at $t=0$ and we set the scale factor $a=1$ initially.
 The resulting distribution for $\vert X_k\vert$ is of Rayleigh type, while the phase of
 $X_k$ has a random uniform distribution. This gives the mean--squared value
 \begin{equation}
\langle{\vert X_k\vert^2} \rangle=\frac{1}{2\omega_k}\,,
 \end{equation}
which corresponds to the vacuum contribution to $\langle h^2_c \rangle$.
Formally, it diverges in the UV and must be regularized. For instance, one can impose a momentum cutoff, compute the Higgs variance as a function of time and subtract the vacuum contribution: $\langle h^2_c \rangle_{\rm reg}= \langle h^2_c \rangle - \langle h^2_c \rangle_{\rm vac}$. This is known as adiabatic regularization (see \cite{Postma:2017hbk} for a recent discussion).

Let us focus for definiteness on the ``standard'' parametric resonance $A=2q$ (fig.~\ref{fig1}).
When the resonance is efficient ($q\gg1$), $ X_k$
for  comoving momenta $k< k_* \sim m q^{1/4}$
grow fast and the mode occupation numbers become large. Away from the inflaton zero crossings, these can be approximated by
 $n_k \simeq \omega_k\vert X_k \vert^2 $.  The modes much above $k_*$ do not get amplified and remain in the ``vacuum state''. Clearly, they do not contribute to $\langle h^2_c \rangle_{\rm reg}$.
 
 In practice, if the momentum cut--off is of order $k_*$, the vacuum subtraction is unimportant since for this  momentum range $\langle h^2_c \rangle$ is dominated by modes with large occupation numbers. In our simulations, the cut--off is taken to be $35m$ which is enough to capture  the UV behavior of the system. The boundary of the stability region in fig.~\ref{fig:final-result} corresponds to substantial occupation numbers up to
 ${\cal O}(100)$ such that the system is in the semi--classical regime $n_k \gg 1$.

We emphasize that   large Higgs variance $\langle h^2_c \rangle_{\rm reg}$ leads to destabilization only if the system is semi--classical, i.e. the occupation numbers are large. For a purely quantum system, large   $\langle h^2_c \rangle_{\rm reg}$ cannot change the ``vacuum state''. Indeed, a single highly energetic particle (or a narrow wave packet)
can lead to      $\sqrt{\langle h^2_c \rangle_{\rm reg}} > h_{\rm crit}$, but that does not lead
to destabilization.
For the same reason, extremely high energy cosmic rays  cannot destabilize the electroweak vacuum \cite{Arnold:1989cb} since a sufficiently large semi--classical patch  of the field in a new phase must be created for the transition.
This is relevant to the weak resonance regime $q\lesssim 1$, where particle creation is inefficient. Even if at late times
  $\sqrt{\langle h^2_c \rangle_{\rm reg}} > h_{\rm crit}$, no transition occurs. We also note that when the resonance has faded away, the dynamics become rather complicated due to rescattering, thermalization and possible non--perturbative phenomena.

\bibliographystyle{JHEP}
\bibliography{bibfile}

\end{document}